\documentclass[prb,floatfix,amsmath,amssymb,superscriptaddress]{revtex4}
\usepackage{graphicx}
\usepackage{latexsym}
\usepackage{graphicx}
\usepackage{times}
\usepackage{color}
\usepackage{amsmath}
\usepackage{dcolumn}
\usepackage{latexsym,amsmath,amssymb,bm,euscript}
\begin{document} 

\title{Universal Resistances of the Quantum RC circuit} 

\author{Christophe Mora}
\affiliation{Laboratoire Pierre Aigrain, \'Ecole Normale Sup\'erieure, Universit\'e Denis Diderot, CNRS; 24 rue Lhomond, 75005 Paris, France}
\author{Karyn Le Hur}
\affiliation{Departments of Physics and Applied Physics, Yale University, New Haven, Connecticut 06520, USA}

\date{\today} 

\begin{abstract} 
\bf
We examine the concept of universal quantized resistance in the AC regime through the fully coherent quantum RC circuit comprising a cavity (dot) capacitively coupled to a gate and connected via a single spin-polarized channel to a reservoir lead. As a result of quantum effects such as the Coulomb interaction in the cavity and global phase coherence, we show that the charge relaxation resistance $R_q$ is identical for weak and large transmissions and it changes from $h/2e^2$ to $h/e^2$  when the frequency (times $\hbar$) exceeds the level spacing of the cavity; $h$ is the Planck constant and $e$ the electron charge. For large cavities, we formulate a correspondence between the charge relaxation resistance $h/e^2$ and the Korringa-Shiba relation of the Kondo model. Furthermore, we introduce a general class of models, for which the charge relaxation resistance is universal. Our results emphasize that the charge relaxation resistance is a key observable to understand the dynamics of strongly correlated systems.
\end{abstract} 

\maketitle 


\noindent

The Landauer-B\" uttiker formula for coherent DC transport lies at the heart of modern electronics \cite{Landauer,Landauer2,Buttiker0} and embodies one of the most dramatic predictions of modern condensed-matter physics: the perfect quantization, in steps of $e^2/h$, of the maximum electrical conductance in one-dimensional metallic channels. It is universal insofar as one may validly neglect the disruptive influences of inelastic scattering processes within the transport process. An elementary explanation of the quantization views the constriction as an electron wave guide which has a non-zero resistance even though there are no impurities, because of the reflections occurring when a small number of propagating modes in the wave guide is matched to a large number of modes in the reservoirs \cite{Imry,Beenakker}. This conductance quantization has been observed in various systems such as quantum Hall states \cite{Klitzing}, quantum point contacts \cite{Wees,Wharam}, carbon nanotubes \cite{Frank,Laughlin} and the helical edge liquid of topological insulators \cite{Konig}. In this manuscript, we thoroughly investigate the AC regime and show that the charge relaxation resistance may be universally quantized; while the quantized resistance in the DC case requires a perfectly transmitted channel \cite{Yacoby,Bachtold}, the charge relaxation resistance remains quantized regardless of the mode transmission. More specifically, we consider  the quantum RC circuit of Fig. 1 in the context of spin-polarized electrons. Theoretically, the study of AC coherent transport was pioneered in a scattering approach by B\" uttiker, Pr\^etre and Thomas~\cite{Thomas2} where a universal charge relaxation resistance of $R_q = h/2 e^2$ was predicted~\cite{Thomas} for a single-mode resistor; the factor $1/2$ is purely of quantum origin and must be distinguished from spin effects. Coulomb blockade effects \cite{Averin,Nazarov} were ignored and later they have been partially included in an Hartree-Fock theory~\cite{Buttiker1}. The quantum mesoscopic RC circuit has been successfully implemented in a two-dimensional electron gas
and the charge relaxation resistance $R_q = h/2 e^2$ was measured \cite{Gabelli,Feve}.
Our work completes the proof of the universal quantized resistance $R_q = h/2 e^2$ by including interactions in the cavity non-perturbatively (exactly). Moreover,  we evidence a mesoscopic crossover at finite frequency $\omega$, where the charge relaxation resistance changes from $h/2e^2$ to $h/e^2$ regardless of the mode transmission. Our findings are obtained in the two complementary limits of weak and large transparencies at the dot-lead interface, and close to the absolute zero in order to preserve the quantum coherence. 

The crossover takes place
when the level spacing of the cavity $\Delta$ becomes equal to $\hbar\omega$; hereafter, we set $\hbar=1$ and $h=2\pi$. For small  cavities and small frequencies, the interacting model results in $R_q = h/2 e^2$. The metallic regime of large cavities~\cite{Berman,Lehnert} is characterized by a continuous spectrum. 
We use a mapping to the charge-Kondo effect~\cite{Matveev1991,Matveev1995,Brouwer,LeHur} to justify the other universal value  $R_q = h/e^2$. Interestingly, this charge relaxation resistance is equivalent to two Sharvin-Imry contact resistances $h/2e^2$ in series; 
\begin{figure}[h]
\includegraphics[width=6.1cm]{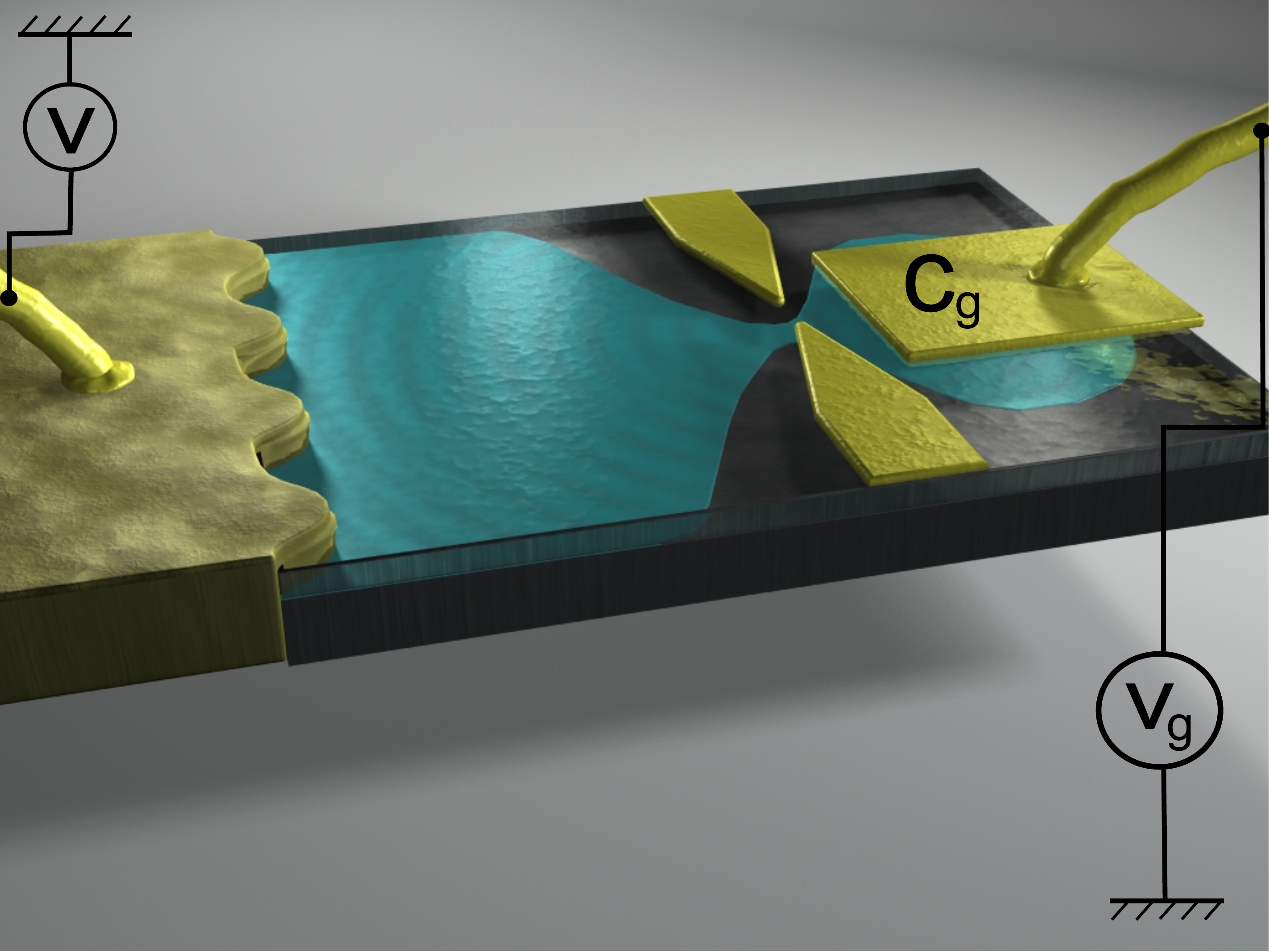}
\hskip 1cm
\includegraphics[width=7.2cm]{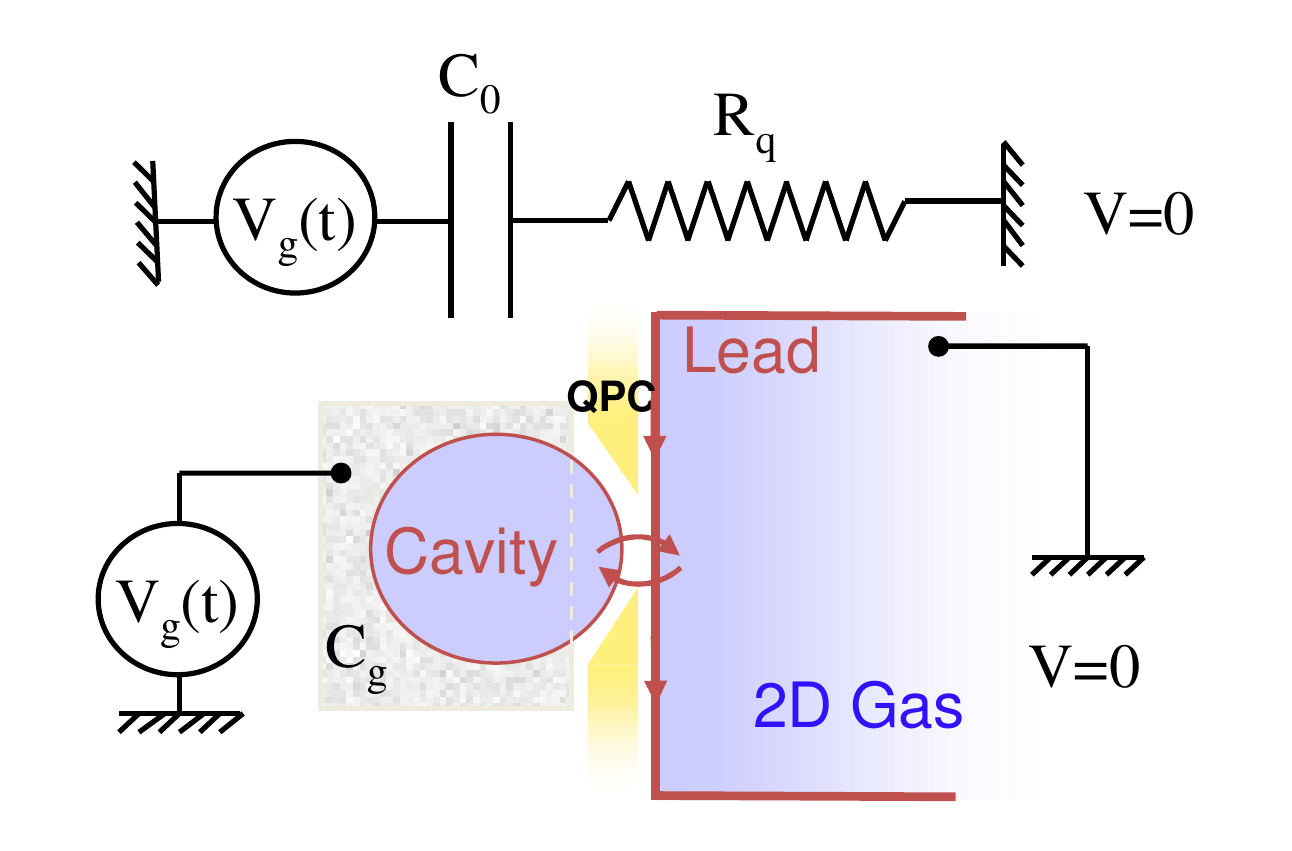}
\caption{The quantum RC circuit realized in a two-dimensional electron gas (2D Gas) \cite{Gabelli} and its equivalent
circuit.}
\end{figure} 
in the metallic regime, an electron entering the cavity is disentangled from an electron escaping the cavity. This result differs from the AC response of a wire coupled to two leads where the contact resistances are added in parallel~\cite{Safi,Burke,blanter,Pham}. 
Recently, the quantum RC circuit has also gained a growing interest in other parametric regimes, both theoretically \cite{Guo,nigg2008,Ioselevich} and experimentally \cite{Delsing}.

\begin{figure}[h]
\includegraphics[width=6.9cm,height=5.3cm]{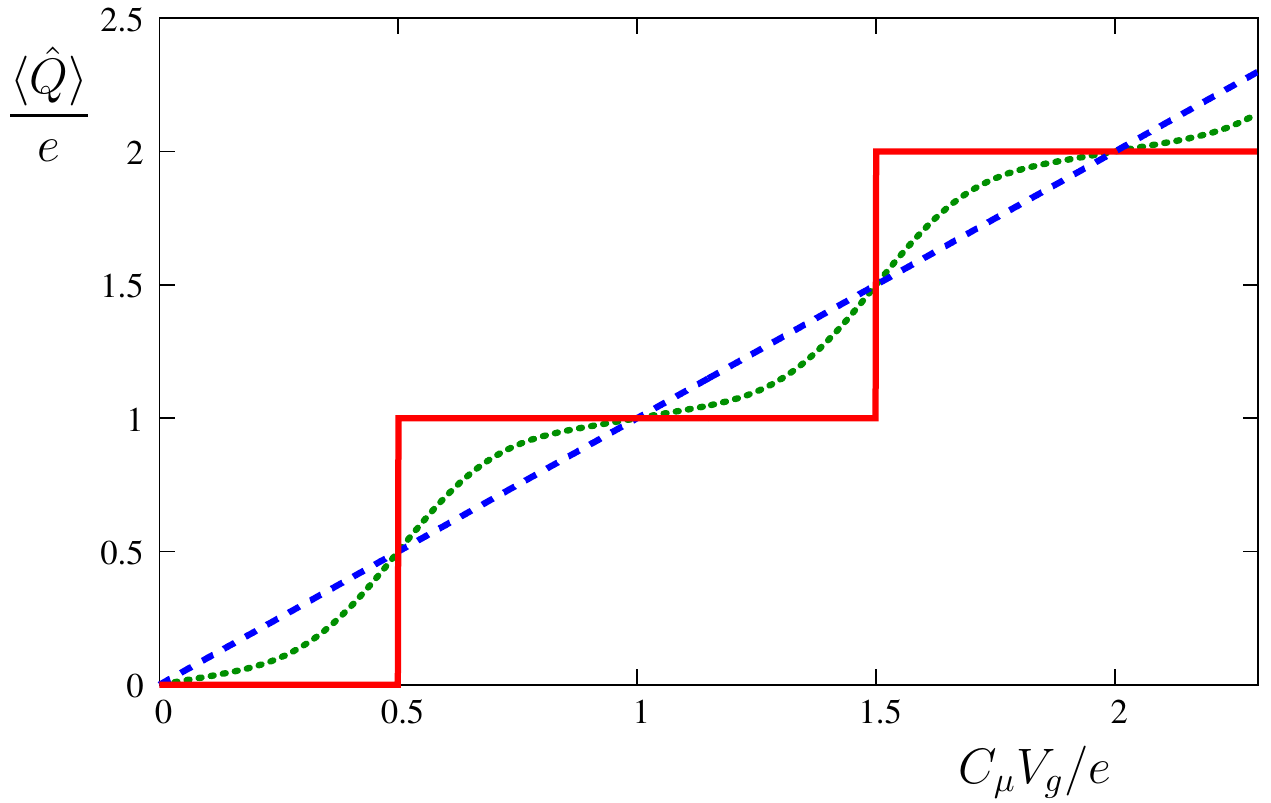}
\caption{Schematic representation of the static charge $\langle \hat{N} \rangle =\langle \hat{Q} \rangle/e$
on the cavity as function of the gate voltage. Different transparencies $D=0$ (plain red), intermediate $D$ (dotted green) and $D=1$ (dashed blue)
are represented. As $D$ increases, charge quantization is reduced. The steps are replaced by oscillations, finally
yielding a linear dependence. The derivatives of these curves give the capacitance $C_0 = \partial \langle \hat{Q} \rangle/\partial V_g$.
It is flat for $D=1$ ($= C_\mu$, see main text) and builds peaks at the charge degeneracy points as the charge becomes quantized.}
\label{charging}
\end{figure}

The potential of the reservoir is taken as a reference $(V=0)$ and we vary the AC potential of the gate $V_g$. The formula that gives the charge in the capacitor at low frequency, 
\begin{equation}\label{charge}
\frac{Q(\omega)}{V_g(\omega)} = C_0 (1 + i \omega C_0 R_q) + {\cal O}(\omega^2),
\end{equation}
 for a classical circuit extends to the quantum regime with modified values of the capacitance $C_0$ and the charge relaxation resistance $R_q$. In particular  phase-coherent transport implies that the capacitive and tunneling effects can not be disentangled. $C_0$ describes the static charging of the dot \cite{Matveev1991,Matveev1995,Grabert}. It is generally different from the geometrical capacitance $C_g$ and depends strongly on the lead-dot transparency $D$ as shown in Fig.~\ref{charging}.
The average of $C_0$ over oscillations as a function of $V_g$ equals the electrochemical capacitance~\cite{Thomas} $C_\mu$, given by the geometrical capacitance $C_g$ in series with the quantum capacitance $e^2/\Delta$.
Similarly $R_q$ does not coincide with the DC resistance $h/(D e^2)$. The reason
for this discrepancy is that carriers injected into the cavity may not equilibrate in AC transport.
The product $R_q C_0$ sets the time scale for the charge to relax and $R_q$ controls energy dissipation during AC driving.
 Below, we provide an exact derivation of $R_q$ and $C_0$ from the weak and large transmission limits.
 
{\bf Weak Transparency.}
As a first example, we consider the case where the cavity is weakly coupled (via tunnel contact)
to the reservoir lead. Matveev has introduced a rigorous formalism allowing to compute the static charge fluctuations on the cavity~\cite{Matveev1991}. We extend the analysis by computing the dynamics of the charge fluctuations at low frequency. The Hamiltonian splits as $H  = H_0  + H_c + H_T$ with
\begin{subequations}\label{hamil}
\begin{align}
\label{h0} H_0 & = \sum_p \varepsilon_p \, c^\dagger_p c_p 
+ \sum_k \varepsilon_k \, d^\dagger_k d_k, \\[1mm]
\label{inter} H_c & = E_c (\hat{N}-N_0)^2, \\[1mm]
\label{tunnel} H_T &= t \sum_{k,p} \left( d^\dagger_k c_{p} + c^\dagger_p d_{k} \right).
\end{align}
\end{subequations}
The single-particle energies $\varepsilon_{k,p}$, in the lead and in the dot are measured from the Fermi level, and $d_k$ and $c_p$ are the corresponding electron annihilation operators. The lead is a metallic reservoir. The level spacing in the dot is finite, $\Delta=\pi v_F/L$ for a 
one-dimensional dot of length $L$. The metallic dot  (large cavity)
limit means $\Delta \to 0$. Here, $E_c =e^2/(2 C_g)$ is the charging energy determined by
the geometrical capacitive coupling to the gate; $\hat{N}$ is the (integer)
operator that gives the number of electrons on the dot and $N_0 = C_g V_g/e$ is
imposed by the gate voltage. Since observables are periodic in $V_g$ ($N_0$),
we restrict $N_0$ to the window $[-C_\mu/(2 C_g),C_\mu/(2 C_g)]$ and charge degeneracy
is reached at the boundaries. The Hamiltonian part $H_T$ describes the 
tunneling of electrons between the lead and the dot. 
Each tunneling event changes the number of electrons on the dot by $\pm 1$.

In the presence of a small time-dependent perturbation of the gate
voltage, the charge on the dot $Q = e \langle \hat{N} \rangle$ obeys,
\begin{equation}\label{linear}
Q (\omega) = e^2 K (\omega) V_g (\omega),
\end{equation}
where the retarded response function, following standard linear response theory
$K ( t - t') = i \theta(t-t')\langle [\hat{N} (t), \hat{N} (t') ] \rangle$, describes charge fluctuations at equilibrium. 
In the absence of electron tunneling, the cavity charge in the ground state is $\langle\hat{N}\rangle = 0$ and does not fluctuate, hence $K = 0$. For weak tunneling, the charge fluctuations on the cavity are determined using perturbation theory in $H_T$.
The second order in the static case ($\omega=0$) gives the capacitance $C_0 = e^2 K(0)$. In the metallic case, one finds \cite{Matveev1991}
\begin{equation}\label{C01}
C_0 = \frac{D C_g}{1/4- N_0^2},
\end{equation}
which diverges a the charge degeneracy points $N_0=\pm 1/2$. The transparency $D = t^2 \nu_0 \nu_1$ depends on the density
of states $\nu_0$ ($\nu_1$) in the lead (cavity). For a finite level spacing $\Delta$, a discrete summation over single-particle
states remains in the expression of $C_0$.
We now turn to the imaginary part of $K(\omega)$ at low frequency describing dissipation. In the presence of a slowly time-varying
gate voltage $V_g (t)$, a redistribution in the virtual occupation of excited states allows the charge $Q$ to follow $V_g$
and creates dissipation~\cite{Glazman}. The only way to dissipate  at small frequency $\omega$ is to find a continuum of excitations with energies
$\sim \omega$ close to the ground state. States with a charge on the cavity different from zero, such as those reached by second 
order perturbation, exhibit gaps $\sim E_c \gg \omega$. Dissipation is thus carried out by states with a charge $e\langle\hat{N}\rangle=0$
but with an additional electron-hole excitation (in the lead or in the dot). This state is reached by applying at least 
two times $H_T$ on the ground state. Hence,  ${\rm Im} \, K (\omega)$ is obtained from the fourth order in perturbation theory.  
For a large cavity where dissipation can occur both in the lead or in the cavity, we find
$ {\rm Im} \, K (\omega ) =  (\pi \omega/2) ( D / E_c )^2 (1/4 - N_0^2)^{-2} $. Comparing with Eq.~\eqref{C01} this results in the illuminating formula,
\begin{equation}\label{reim}
{\rm Im} \, K (\omega ) =  2 \pi  \omega \left[ {\rm Re} K (0 ) \right]^2,
\end{equation}
at small frequency $\omega$. The same calculation is carried out for a small cavity and in this case we find:
\begin{equation}\label{reim2}
{\rm Im} \, K (\omega ) =  \pi  \omega \left[ {\rm Re} K (0 ) \right]^2.
\end{equation}
Physically, states in a small cavity do not form a continuum (since $\Delta > \omega$) and therefore they do not contribute to
dissipation at small frequencies. This explains the reduction by a factor $1/2$ between Eq.~\eqref{reim2}
and Eq.~\eqref{reim}. The charge response to a gate voltage oscillation in Eq.~\eqref{linear} is matched with the RC circuit formula~\eqref{charge}. Mostly, Eq.~\eqref{reim} gives $R_q = 2 \pi/e^2 = h/e^2$ for large cavities and Eq.~\eqref{reim2} gives $R_q = \pi/e^2 = h/ 2e^2$ for small cavities. From the above discussion, it appears that the crossover between the two regimes of cavity takes place when $\omega \simeq \Delta$ where the states in the cavity merge into a continuum able to dissipate energy.
\\

{\bf Charge relaxation resistance and Korringa-Shiba formula.} In fact, the result of Eq.~\eqref{reim} for large cavities can be inferred from the charge degeneracy point $N_0 = 1/2$ where perturbation theory breaks down and Eq.~\eqref{C01} for $C_0$ is seen to diverge.
In the vicinity of $N_0= 1/2$, the states with a charge $0$ and $e$ on the cavity give a resonant and dominant contribution
to the charge fluctuations. It has been realized by Matveev~\cite{Matveev1991} that, by removing other charge states, a
mapping to the anisotropic Kondo model can be formulated. The fictitious spin ${\bf S}$ of the Kondo model 
acts on the space formed by the charge state $\langle \hat N\rangle=0$ and $1$ with
\begin{equation}\label{nsz}
\hat{N} = \frac{1}{2} - S^z.
\end{equation}
The spin up and down conduction electrons in the Kondo model originate from cavity and lead electrons respectively,
whose Hamiltonian is given by  Eq.~\eqref{h0}.
The tunneling term of Eq.~\eqref{tunnel} changes the charge state corresponding to a spin-flip event. Hence $2 t$ plays the role of
the antiferromagnetic (transverse) Kondo coupling. Finally, the vicinity to the charge degeneracy $E_c (1-2 N_0)$ gives a local magnetic field coupled
to $S^z$.
The correspondence~\eqref{nsz} between charge and spin indicates that the charge fluctuations
$K (\omega)$ are related to the longitudinal spin susceptibility in the effective Kondo model,
\begin{equation}
K ( t - t') = \chi_{zz} (t-t') = i \theta(t-t')
\langle \left[ S^z (t), S^z (t') \right] \rangle.
\end{equation}
The anisotropic and isotropic Kondo models converge to the same Fermi liquid fixed point \cite{Nozieres}.
Thus, at zero temperature, one may use the Korringa-Shiba
relation \cite{Shiba} at low frequency $\omega$
\begin{equation}\label{korrshi}
{\rm Im} \, \chi_{zz} (\omega ) =  2 \pi  \omega \left[ {\rm Re} \chi_{zz} (0 ) \right]^2,
\end{equation}
to recover  Eq.~\eqref{reim} and therefore the universal charge relaxation resistance $R_q =  h/e^2$.
The demonstration of Eq.~\eqref{korrshi} in Ref.~\onlinecite{Glazman} is quite instructive. It shows that a single term accounts for both the static susceptibility and dissipation explaining the structure of Eq.~\eqref{korrshi}.
Moreover this demonstration indicates that, similarly to the perturbative region (far from $N_0=1/2$),
dissipation is caused by electron-hole excitations, yielding the linear low frequency dependence for ${\rm Im} \, \chi_{zz}(\omega)$.
The conclusion is that the Korringa-Shiba relation~\eqref{korrshi}  generally applies at a Fermi liquid fixed point and in particular in the context of the anisotropic Kondo model \cite{Slezak} at low temperatures and in the presence of any magnetic field \cite{Glazman}. We recall that a magnetic field in the spin model corresponds to the distance $E_c (1-2N_0)$ to the charge degeneracy
point in the original charge model.
Eq.~\eqref{reim} and the charge relaxation result $R_q =  h/e^2$ therefore apply for all values of $N_0$,
from the vicinity of $N_0 = 1/2$ to the perturbative region where these two formulas have been checked explicitely.
An intimate connection between the  Korringa-Shiba in its generalized form~\cite{Glazman} and the universal resistance
 $R_q =  h/e^2$ is revealed here, where the Fermi liquid nature of low energy particle-hole excitations plays a striking
role. The generality of these arguments in fact suggests that the result $R_q =  h/e^2$ (and $R_q =  h/2 e^2$) are indeed
universal in the sense that they do not depend either on the gate voltage $N_0$ or on the transparency $D$.
\\

{\bf Large Transparency.} We pursue this idea and investigate the opposite limit at and close to perfect transmission. In what follows, we model
the complete system by electrons moving along a one-dimensional line, the lead is between $-\infty$ and $-L$ and the cavity
between $-L$ and $0$. The level spacing on the isolated cavity is still $\Delta = \pi v_F/L$.
The interaction term~\eqref{inter}, and in particular the Coulomb blockade phenomenon in the cavity, can be treated
exactly using the bosonization approach \cite{Gogolin,Giamarchi}. Integrating all irrelevant modes in an action formalism, one finds,
at perfect transmission $D=1$, the action
\begin{equation}\label{actionS0}
S_0 = \frac{1}{\pi} \sum_n \phi_0 (\omega_n) \phi_0 (-\omega_n)
\left[ \frac{|\omega_n|}{1 - e^{-2| \omega_n| L / v_F}}  + \frac{E_c}{\pi} \right],
\end{equation}
where $\omega_n = 2 \pi T n$ denote bosonic Matsubara frequencies; the Boltzmann constant $k_B$ is set to unity. Here, the field $\phi_0$ is related to the charge 
on the cavity, $\hat{N} = \tilde{N}_0 + \phi_0/\pi$ with 
$\tilde{N}_0 = C_\mu N_0/C_g=C_\mu V_g/e$. From this quadratic action, the response 
function~\eqref{linear} is straightforwardly calculated. We find:
\begin{equation}\label{charge2}
\frac{Q (\omega)}{V_g (\omega)} = \frac{C_g }{ 1  - \frac{i \omega \pi /E_c}{1-e^{2 i \pi \omega /\Delta}}}.
\end{equation}
Interestingly, dissipation vanishes and the model becomes purely capacitive~\cite{blanter} each time the frequency $\omega$ hits a multiple of $\Delta$ corresponding to an eigenstate of the isolated cavity.
At low frequency $\omega\ll \Delta$, we extract $C_0 = C_\mu$ - meaning that the Coulomb blockade effect vanishes~\cite{Matveev1995}
 for perfect tranparency $D=1$ - and $R_q = h/ 2e^2$ from the comparison of Eq.~\eqref{charge2} to the classical RC circuit
formula~\eqref{charge}. We now discuss the transition to large metallic cavities. Eq.~\eqref{charge2}  shows an oscillatory behavior for $\omega > \Delta$.
We thus average over a finite bandwidth $\delta \omega$, such that $\omega \gg \delta \omega \gg \Delta$, and finally we find:
\begin{equation}\label{charge3}
\frac{Q (\omega)}{V_g (\omega)} = \frac{C_g }{ 1  - i \omega \pi /E_c}.
\end{equation}
This result is also obtained if one takes $L \to +\infty$ in Eq.~\eqref{actionS0}.
It can be checked that this correspondence extends to all correlation functions. Hence, Eq.~\eqref{actionS0}
with $L \to +\infty$ defines the action for the large cavity regime, as shown in the Methods.
A comparison of the result~\eqref{charge3} with Eq.~\eqref{charge} gives
$C_0 = C_\mu = C_g$ and $R_q = h/e^2$. We thus recover the universal resistances $h/2 e^2$ and $h/e^2$ for the small and large cavities, and the fact that the crossover takes place when the frequency $\omega$ becomes larger than $\Delta$.

Backscattering at the interface between the cavity and the lead (at $x=-L$) may be incorporated in the model as,
\begin{equation}\label{backsca}
S_{BS} = - \frac{v_F r}{\pi a} \int_0^\beta d \tau \cos \left[ 2 \phi_0(\tau) + 2 \pi \tilde{N}_0 \right],
\end{equation}
and the total action is now given by $S = S_0 + S_{BS}$ with Eq.~\eqref{actionS0} and Eq.~\eqref{backsca}.
Here, $a$ is a standard short-distance cutoff and $1/a$ defines the region around $k_F$
where the electron band spectrum can be linearized. Here, $r$ is the dimensionless strength of backscattering
and the model only involves a single phase $\phi_0$; see Methods. For large $r$, or small transparency $D$, $\phi_0$ gets frozen around the values which minimize Eq.~\eqref{backsca}.
Translated into the charge of the cavity, this gives $\langle\hat{N}\rangle = n \in {\bm N}$ and we recover charge quantization
as shown in Fig.~\ref{charging}. The non-linearity of Eq.~\eqref{backsca} does not allow a complete analytical approach.
We thus consider the case of weak backscattering at the interface, where $r \ll 1$ and the transparency is given by $D = 1-r^2$.

Let us first discuss the calculation of the charge relaxation resistance $R_q$.
In order to remain quite general in the following discussion, we define 
the imaginary time non-interacting Green's function
$G_0 (\tau)= \langle \phi_0 (\tau)
\phi_0(0) \rangle_0$ where the mean value $\langle \ldots \rangle_0$ is taken with respect  to
the quadratic action $S_0$
given  by Eq.~\eqref{actionS0}. At low (real) frequency, we expand  to first order 
\begin{equation}
G_0 (\omega) = \frac{\pi^2}{2 E_c} \left( A + i \frac{\omega \pi}{E_c} B \right),
\end{equation}
and the values of $A$ and $B$ depend on whether  a small or large cavity is considered: $A^{-1} =1+\Delta/2 E_c$,
$B^{-1} = 2 (1+\Delta/2 E_c)^2$ and, $A= 1$, $B=1$, respectively. The calculation of $R_q$ goes as follows:
 the charge fluctuations $K(\omega)$ (see Eq.~\eqref{linear}) are computed perturbatively to second order in $r$ and 
 the result is expanded to first order in $\omega$. The identification with Eq.~\eqref{charge} leads to
\begin{equation}\label{rquni}
R_q = \frac{h}{e^2} \, \frac{B}{A^2},
\end{equation}
and the universal charge relaxation resistances $h/2 e^2$ and $h/e^2$ for the small and large cavities are recovered.

The capacitance $C_0$ is determined from the static charge fluctuations ($\omega =0$).
In contrast with the charge relaxation resistance, the result is non-universal and a continuous function of the ratio $\Delta/E_c$ describing a crossover between a weakly ($E_c \ll \Delta$) and a strongly ($\Delta \ll E_c$) interacting regime.
It takes the form
\begin{equation}\label{resultcapa}
\frac{C_0}{C_\mu} = 1- \frac{r e^C g_1 (\Delta/E_c) }{\pi} 2 \pi
 \cos (2 \pi \tilde{N}_0) + \left( \frac{r e^C g_1 (\Delta/E_c)}{\pi} 
 \right)^2 g_2 (\Delta/E_c) 4 \pi \cos(4 \pi \tilde{N}_0),
\end{equation}
where $C = 0.577\ldots$ is the Euler constant. The function $g_1$ smoothly connects $g_1(0) = 1$ to $g_1 (\infty)=e^{-C}$, and
$g_2$ connects $g_2(0) \simeq 1.86$ to  $g_2(\infty) = \pi/2$. Our results for the capacitance $C_0$ are in agreement with calculations  in the metallic limit~\cite{Matveev1995,Clerk} for $\Delta \ll E_c$,
and with calculations based on the non-interacting AC scattering 
approach~\cite{Thomas2,Gabelli} for $E_c \ll \Delta$.
\\

{\bf Discussion and Outlook.} It is important to stress that Eq.~\eqref{rquni} is not a trivial result. Instead it is the result of a remarkable cancellation of all $r^2$ corrections, which protects the value of $R_q$. This cancellation occurs for all (positive) values of
$A$ and $B$ and takes roots in the structure of the backscattering term~\eqref{backsca}.
It is tempting to conjecture, based on our previous results in the weak-tunneling regime, that this cancellation is present at all orders in $r$, {\it i.e.}, for all transparencies. We also envision a whole class of models, parametrized  by the values of $A$ and $B$, such that
the value of the charge relaxation resistance is universal and simply given by Eq.~\eqref{rquni}. 
For example, our analysis extends to the fractional quantum Hall regime where the edge state, lead and cavity,
 embodies   a chiral Luttinger liquid~\cite{Wen} described through the same bosonization framework~\cite{Giamarchi}.
We shall discuss it here  briefly.
The action can be written $S_0'+S_{BS}$ where $S_{BS}$ is exactly Eq.~\eqref{backsca} and $S_0'$ has the
same structure as Eq.~\eqref{actionS0} but different coefficients. In particular, 
$A$ is divided by $g^2$, $B$ by $g^3$ and the cavity charge is multiplied by  $g$,  
where $g = 1/q$, with $q$ odd, is the bulk filling factor $\nu$.
For $g=1$,  $S_0'=S_0$ and the  locally noninteracting case is recovered.
The same analysis can be repeated and one finds the universal charge relaxation resistances $R_q = h/(2 g e^2)$ and $R_q = h/(g e^2)$ for small and large cavities.

It is not surprising, in a sense, to recover the resistance $h/e^2$ for a large cavity in the large transparency limit. Matveev has argued in detail~\cite{Matveev1995} that the large transmission regime describes the strong-coupling fixed
point of the fictitious Kondo model that we already discussed for small transparencies. The Korringa-Shiba relation~\eqref{korrshi}
(or Eq.~\eqref{reim}) should therefore also apply for $D \simeq 1$, leading to $R_q = h/e^2$.
In fact, in the same way Matveev argues that the logarithmic singularities of  the capacitance $C_0$ in the case of spin-unpolarized electrons
should appear for all transparencies, one could argue that it is also the case for the Korringa-Shiba relation, and $R_q = h/e^2$. Based on these renormalization group arguments, we spectulate that the Korringa-Shiba relation influences the whole phase diagram and emerges as the fundamental reason for a universal $R_q$. 
A complementary interpretation of our results is that the charge relaxation for coherent transport in a small cavity equals the contact resistance $h/2 e^2$,
regardless of the mode transmission. For large cavities, electrons entering the cavity and electrons escaping the cavity are uncorrelated
and the charge relaxation resistance thus coincides with two contact resistances in series.
 Finally, it should be noted that our results are not limited to the close vicinity of zero temperature; in particular, for large transparencies, we find that the two universal charge relaxation resistances remain unchanged for temperatures  comparable to $E_c$.
\\

{\bf Summary.} In this manuscript we have developed a general, unifying framework by investigating the relation between dissipation and resistance in the quantum RC circuit. Focusing on the `interactions with coherence' regime
close to the absolute zero, we have demonstrated that the charge relaxation resistance of the fully coherent
quantum RC circuit is universal regardless of the single spin-polarized mode transmission, and that the universal quantized value switches from $h/2e^2$ to $h/e^2$ when the frequency (times $\hbar$) exceeds the level spacing of the cavity. This emphasizes that in the quasi-static limit of low frequencies, quantum effects such as electron coherence and interaction already invalidate the Kirchhoff's law, which relates the resistance of a classical system to the tunneling conductance of the contact. Our framework gives a new perspective on how to broadly discuss the dynamics of strongly correlated systems through the concept of charge relaxation resistance. In particular, in a simple and intuitive picture, the universal quantized value $h/e^2$ 
has been related to the strong-coupling fixed point of the Kondo model. 

{\it Acknowledgments.---} We thank M. B\" uttiker, G. F\` eve, C. Glattli, T. Martin, and B. Pla\c{c}ais for stimulating discussions. K.L.H. was supported by Department of Energy, under the grant DE-FG02-08ER46541. This work has applications for the manipulation
of quantum systems and therefore K.L.H. also thanks the Yale Center
for Quantum Information Physics (NSF DMR-0653377).

\section*{\large Methods}

{\bf Open boundary bosonization.} The interacting model at and close to perfect transmission can be solved exactly
by applying the open boundary bosonization framework \cite{Gogolin}. Here, the whole system occupies a half-infinite
one-dimensional line that stops at $x=0$, with only negative values of $x$. The dot corresponds to the region between $x=-L$ and $x=0$; backscattering occurs at the entrance of the dot at $x=-L$. One-dimensional
fermionic fields are usually decomposed in terms of left and right moving fields, $\psi_L(x)$ and $\psi_R(x)$ respectively. The idea behind open boundary bosonization is that the semi-infinite line is unfolded such that $L$-moving electrons become $R$-moving electrons on the positive $x$ axis, namely, $\psi_L ( x ) = - \psi_R (-x)$. We are left with a chiral infinite line of fermions. Then, the right-moving field can be bosonized using
a single boson field, $\psi_R (x) = e^{i \phi(x)}/\sqrt{2\pi a}$, where $a$ is a short-distance cutoff. Within these
notations, the total Hamiltonian takes the form $H=H_0+H_{BS}$, where
\begin{subequations}\label{fullh}
\begin{align}
\label{h02} H_0  & = \frac{v_F}{4 \pi} \int_{-\infty}^{+\infty} d x \left[  \partial_x \phi (x) \right]^2 + \frac{E_c}{\pi^2} \left( \frac{\phi(L)-\phi(-L)}{2} -\pi N(t) \right)^2 \\[1mm]
H_{BS} & = - \frac{v_F r}{\pi a} \cos \left[ \phi(L)-\phi(-L) \right].
\end{align}
\end{subequations}
The first term in $H_0$ embodies the kinetic term and $v_F$ is the Fermi velocity. The second term containing $E_c = e^2/2 C_g$ represents the charging energy (the interaction on the cavity) and $e N(t) = C_g V_g (t)$. The charge on the cavity takes the form $e \int_{-L}^{L} d x  \, \psi_R^\dagger (x)   \psi_R (x)=e(\phi(L)-\phi(-L))/(2\pi)$. In addition, $H_{BS}=- v_F r \left( \psi_R^\dagger (-L) \psi_R (L) + {\rm h.c.} \right)$ describes the backscattering of electrons at the entrance of the cavity at $x=-L$ and $r$ depicts the dimensionless backscattering strength. The strategy then is to integrate out the irrelevant
modes different from $\phi_0=\frac{\phi(L)-\phi(-L)}{2}$ in the kinetic term. Following a standard procedure \cite{Giamarchi}, this results in Eq. (\ref{actionS0}).
We consider small oscillations of the gate voltage around some constant mean value such that $N(t)=N_0+N_1(t)$ with $|N_1(t)|\ll N_0$. Therefore, it is convenient to shift the field $\phi_0$ by the constant $\pi \tilde{N}_0$, leading to Eq. (\ref{backsca}). Linear response theory relates the charge on the dot to the gate voltage,
$Q (\omega) = e^2 K (\omega) V_g (\omega)$. The correlation function $K (\tau) = \frac{1}{\pi^2}  \langle T_\tau \phi_0 (\tau)\phi_0 (0) \rangle$
is Fourier transformed and analytically continued, $i \omega_n \to \omega+i 0^+$,
to produce $K (\omega)$.
\\

{\bf Infinite metallic cavity.} In the limit of a very large cavity \cite{Matveev1995,Brouwer} $L\rightarrow +\infty$, one can assume that electrons entering and electrons escaping the cavity are uncorrelated. For convenience, we slightly change
of conventions; now the cavity lies between $x=0$ and $L\rightarrow +\infty$ and the reservoir lead occupies the negative
$x$ values. In this case, we rather obtain:
\begin{subequations}
\begin{align}
H_0 & = \frac{v_F}{2 \pi} \int_{-\infty}^{+\infty} d x\ 
\pi^2 [ \Pi(x) ]^2  + [ \partial_x \phi (x) ]^2 +  \frac{e^2}{2 C_g} \left[ \frac{\phi(0)}{\pi} - N (t) \right]^2  \\[1mm]
H_{BS} & = - \frac{v_F r}{\pi a} \, \cos \left[ 2 \, \phi(0) \right].
\end{align}
\end{subequations}
Here, $\Pi (x)= \frac{1}{\pi} \partial_x \theta (x)$ is the 
momentum conjugate to the charge field $\phi (x)$, $[ \phi (x), \Pi (y) ] = i \delta (x -y)$. In the limit of very large cavities, electrons at $x=L\rightarrow +\infty$ completely decouple from the backscattering events at $x=0$ and therefore we may set $\phi(+\infty)=0$. The charge operator on the cavity then reads $e\phi(0)/\pi$. It is certainly appropriate
to notice the analogy between the mode $\phi_0$ introduced above and $\phi(0)$. More precisely, in the case of an infinite cavity, we find
the local action:
\begin{equation}
S_0   = \frac{1}{\pi} \sum_n \phi (0,\omega_n) \phi(0,-\omega_n) \left[
|\omega_n| + \frac{E_c}{\pi} \right],
\end{equation}
which reproduces Eq. (\ref{actionS0}) when $L\rightarrow +\infty$,
allowing us to already justify Eq. (\ref{charge3}) at perfect transmission $(r=0)$. 
\\

{\bf Mesoscopic crossover.} In fact, it is possible to recover the infinite (metallic) cavity regime from the finite dot situation by averaging over a finite bandwidth $\delta\omega$ such that $\omega\gg \delta \omega \gg \Delta$. For $r=0$,  the response function $K(\omega)$ averaged over
$\delta\omega$ reads:
\begin{equation}
K (\omega) = \frac{1}{2 E_c} \, 
\int_0^{2 \pi} \frac{d \varphi}{2 \pi} \,
\frac{1}{ 1  - \frac{i \omega \pi /E_c}{1-e^{i \varphi}}}.
\end{equation}
The integral is computed by changing $z = e^{i \varphi}$, then reproducing the infinite cavity result $K(\omega)=(1/2 E_c)(1-i\omega\pi/E_c)^{-1}$. In fact, this smearing procedure transforms
each term of the perturbative expansion in $r$ to its equivalent in 
the metallic case.


\newpage

\begin{center}
{\large  {\bf Supplementary Material}}
\end{center} 
\section{Perturbative calculation for weak transparency}

Here, we discuss details of the calculation in the regime of weak tunneling between
the lead and the cavity.  The response function $K (\omega)$  
relates the charge on the cavity to the gate voltage,  see Eq.~\eqref{linear} in the main text.
It is given by
\begin{equation}
K (\omega ) = i \int_{-\infty}^{+\infty} d t \, e^{i \omega t} \langle
T_t [\hat{N} (t) \hat{N}(0) ] \rangle,
\end{equation}
where $T_t$ denotes time ordering and $\omega$ has a small positive imaginary part.
The ground state $| \phi_{GS}\rangle$ 
of the Hamiltonian $H_0 + H_c$, (Eqs.~\eqref{hamil} in the main text),
is formed by two filled Fermi seas so that all states with negative 
(positive) energy are filled (empty); the ground state corresponds by definition to a charge $0$
on the cavity. $H_T$ describes the tunneling of electrons and induces 
charge fluctuations on the cavity. It shall be treated using perturbation theory.
At zero temperature, we can write
\begin{equation}\label{correl}
 \langle
T_t [\hat{N} (t) \hat{N}(0) ] \rangle = \frac{\langle \phi_{GS} |
T_t [\hat{N} (t) \hat{N}(0) U(\infty,-\infty) ] | \phi_{GS}\rangle}
{\langle \phi_{GS} |
T_t [ U(\infty,-\infty) ] | \phi_{GS}\rangle},
\end{equation}
and expand the evolution operator
\begin{equation}
 U(\infty,-\infty) = \sum_{n=0}^{+\infty} \left ( -i \right)^n
\int_{-\infty}^{+\infty} dt_1 \int_{-\infty}^{t_1} dt_2 \ldots \int_{-\infty}^{t_{n-1}} dt_n
\, H_T(t_1) H_T (t_2) \ldots H_T(t_n),
\end{equation}
in powers of $H_T (t) = e^{i (H_0+H_c) t} H_T  e^{-i (H_0+H_c) t}$. 
Let us first consider second order perturbation theory. $H_T$ transfers one electron 
from the cavity to the lead or vice versa.  The intermediate states have 
therefore a hole (electron) in the lead. 
Due to the  Coulomb repulsion, these states are well 
separated in energy from the ground state and only virtually occupied.
After Fourier transformation, the response function reads
\begin{equation}\label{secondorder}
K (\omega )  = t^2 \nu_0 \int d \varepsilon \sum_{\varepsilon'} \sum_{\eta_{1/2}=\pm} 
\frac{\theta(- \eta_1 \varepsilon) \theta(\eta_1 \varepsilon')}
{[E_{\eta_1} + \eta_1 (\varepsilon' - \varepsilon)]^2}
\, \frac{1}{ E_{\eta_1} + \eta_1 (\varepsilon' - \varepsilon) + \eta_2 \omega},
\end{equation}
where $E_{\pm} = E_c (1 \mp 2 N_0)$ is the Coulomb gap for the states with charge $\pm e$.
$\theta (\varepsilon) $ denotes the Heavyside function. 
$\varepsilon$ ($\varepsilon'$) corresponds to single-particle energies in the lead (cavity)
and $\nu_0$ is the density of states in the lead at the Fermi level.
A large metallic cavity allows to perform both energy summations. The result at zero frequency
for the capacitance $C_0$ is given by Eq.~\eqref{C01} in the main text.
At finite but small frequency $\omega \ll E_c$, Eq.~\eqref{secondorder} shows no singularity
when the frequency $\omega$ crosses the real axis, hence the imaginary part of $K(\omega)$ vanishes.
Poles are crossed in Eq.~\eqref{secondorder} only for frequencies above the Coulomb gaps
$E_{\pm}$ where the charged states are able to dissipate energy. A related discussion can be found in
the main text.

The leading contribution to ${\rm Im} K(\omega)$ at low frequency is thus obtained by
pushing the perturbation theory to fourth order in $H_T$. A class of intermediate states 
with charge $0$ and one electron-hole excitation on top of the ground state can 
be reached. It forms a continuum with energies arbitrarily close to the ground state and
therefore allows dissipation. The full fourth order in perturbation theory is not needed
since we seek  only the dominant term in ${\rm Im} K(\omega)$ at low frequency.
For $t>0$, the relevant fourth order correction to Eq.~\eqref{correl} is
\begin{equation}\label{fourthorder}
 \int_{t}^{+\infty} dt_1
\int_{0}^{t} dt_2 \int_{0}^{t_2} dt_3
\int_{-\infty}^{0}dt_4
\langle \phi_{GS} | H_T(t_1) \hat{N} (t) H_T (t_2)  H_T (t_3) \hat{N}(0) H_T (t_4) | \phi_{GS}\rangle,
\end{equation}
and the positions of $\hat{N} (t)$ and $\hat{N}(0)$  are inverted when $t<0$.
The charge on the cavity for the second intermediate states should be zero.
Other choices of positions for $\hat{N} (t)$ and  $\hat{N}(0)$  can be checked to lead to vanishing
expressions at low frequency.
Eq.~\eqref{fourthorder} implies fourth possible paths for the cavity charge, with a charge of $\pm e$
for both the first and the third intermediate states.
Moreover, for each of these paths, the electron-hole excitations of the second intermediate states
may end up either in the cavity or in the lead.
We take $\omega >0$ below. 
Let us first discuss the case where the electron-hole is in the lead and the charge follows the path
 $0:1:0:1:0$.  After Fourier transformation, the following contribution is obtained, 
\begin{equation}\label{fourthorder1}
t^4 \nu_0^2 \int d \varepsilon_2 d \varepsilon_3 \sum_{\varepsilon_1,\varepsilon_4} 
\frac{\theta(\varepsilon_1) 
\theta(-\varepsilon_2)
\theta(\varepsilon_3) \theta(\varepsilon_4)}{  (E_+ + \varepsilon_1 - \varepsilon_2 )
(E_+ + \varepsilon_1 - \varepsilon_2 - \omega ) ( E_+ + \varepsilon_4-\varepsilon_2)
 (E_+ + \varepsilon_4-\varepsilon_2 - \omega ) (\varepsilon_3 -\varepsilon_2-\omega)},
\end{equation}
The only relevant $\omega$-dependence in Eq.~\eqref{fourthorder1} (at low frequency) is in the last term of
the denominator and corresponds to the electron-hole excitation. It has a pole, and
a contribution to the imaginary part, for $\varepsilon_3 -\varepsilon_2 = \omega$.
Discarding all other $\omega$-dependences in Eq.~\eqref{fourthorder1}, we integrate over the
leads energies, take the imaginary part and find 
\begin{equation}
 t^4 \nu_0^2 \, \pi \omega \,\left( \sum_{k_1} \frac{\theta(\varepsilon_{k_1})}
{ [ E_+ + \varepsilon_{k_1} ]^2} \right)^2.
\end{equation}
Adding all other paths where the intermediate state has an electron-hole excitation in the lead finally
results in
\begin{equation}\label{smallcav}
{\rm Im} \, K (\omega ) = \pi t^4 \nu_0^2 \, \pi \omega \, \left( \sum_{\varepsilon'} \sum_{\eta=\pm} 
\frac{\theta(\eta \varepsilon')}{ [ E_{\eta} + \eta_1 \varepsilon_{k_1} ]^2} \right)^2,
\end{equation}
for a small cavity with a finite level spacing $\Delta$. The  comparison of Eq.~\eqref{smallcav}
with Eq.~\eqref{secondorder} gives us the Eq.~\eqref{reim2} in the main text.

For a large metallic cavity, the electron-hole excitations of the second intermediate states
in the cavity should also be taken into account. The calculation is similar as in the lead and
multiply Eq.~\eqref{smallcav} by a factor two, hence   Eq.~\eqref{reim} in the main text.

\section{Bosonization at large transmission}

\subsection{Starting model}

Here, we flesh out the concept of open boundary bosonization~\cite{Gogolin} complementing the discussion in the Methods.
The whole system, lead and cavity, occupies a semi-infinite one-dimensional line stopping at
$x=0$. Electrons are described by fermions with two chiralities, as the two points of the Fermi surface,
in one dimension. Alternatively, the boundary condition on the electron field operator $\psi (x=0) = 0$
can be used to unfold the semi-infinite line onto a really infinite one-dimensional line
with fermions that now possess a single chirality. After this transformation, the cavity is defined
between $x=-L$ and $x=L$, and the lead occcupies the rest of the one-dimensional line, $|x| >L$.
Backscattering at the lead-cavity interface couples electrons at $x=-L$ and $x=L$.
The unfolded geometry is in fact the one of an edge state in the Quantum Hall Effect where the two
points $x=-L$ and $x=L$ join to form a cavity of size $2 L$. In particular, this is the geometry of 
Refs.~\onlinecite{Gabelli} and \onlinecite{Feve}. 

The Hamiltonian is written in terms of the chiral electron field operator $\psi_R (x)$,
$H = H_0 + H_{BS}$ with
\begin{subequations}
\begin{align}
H_0  & =  \int_{-\infty}^{+\infty} d x v_F\left[ \psi_R^\dagger (x) 
( - i \partial_x)  \psi_R (x)  \right] +  E_c (\hat{N}-N_0)^2\\[1mm]
H_{BS} & = - v_F r \left( \psi_R^\dagger (-L) \psi_R (L) + {\rm h.c.} \right),
\end{align}
\end{subequations}
where $H_0$ now contains the kinetic energy and the Coulomb interaction in the cavity with capacitive coupling to the gate, and $H_{BS}$ describes backscattering at
the interface. Within these notations, $N_0 = C_g V_g/e$ and the number of electrons in the dot 
is $\hat{N} = \int_{-L}^{L} d x  \, \psi_R^\dagger (x)   \psi_R (x)$.
We proceed and bosonize the chiral electron field operator $\psi_R (x)$ using a single boson
field $\phi(x)$,
\begin{equation}
\psi_R (x) = \frac{1}{\sqrt{2 \pi a}} e^{i \phi(x)},
\end{equation}
 where $a$ is a short-distance cutoff. The electron density reads 
$\psi_R^\dagger (x)   \psi_R (x) = \partial_x \phi (x)/(2 \pi)$ and the charge in the cavity
\begin{equation}
\hat{N} = \frac{\phi(L)-\phi(-L)}{2 \pi}.
\end{equation}
The dynamics then is governed by the Euclidian action, $S = S_0 + S_{BS}$, with
\begin{subequations}
\begin{align}
S_0 & = \frac{1}{4 \pi} \int_{-\infty}^{+\infty} d x \int_0^\beta 
d \tau \, \left[  i (\partial_\tau \phi) (\partial_x \phi)   
+ v_F (\partial_x \phi)^2 \right] +  
E_c \int_0^\beta d \tau \, \left( \frac{\phi(L)-\phi(-L)}{2 \pi} - N_0 \right)^2 \\[1mm]
S_{BS} & = - \frac{v_F r}{\pi a} \int_0^\beta d \tau \cos \left[\phi(L)-\phi(-L) \right].
\end{align}
\end{subequations}
It is convenient to singularize the mode $\phi_0 = \frac{\phi(L)-\phi(-L)}{2}$
which is proportional to the charge in the cavity, $\hat{N} = \phi_0/\pi$, as it also appears
in the backscattering term. Therefore, we integrate all other bosonic modes in the action formalism~\cite{Giamarchi}
and obtain the two terms $S_0$ and $S_{BS}$ given by Eqs.~\eqref{actionS0} and~\eqref{backsca}
in the main text. A constant shift $\phi_0 \to \phi_0 + \pi \tilde{N}_0$, with 
$\tilde{N}_0 = C_\mu N_0/C_g = C_\mu V_g/e$, has also been applied  in order to transfer 
the gate voltage dependence inside the backscattering action.
The large cavity regime is obtained by taking $L \to +\infty$ directly in the action of Eq.~(\ref{actionS0}), as discussed in the Methods.

\subsection{Expansion in the backscattering}

We now detail the perturbative calculation in the backscattering strength $r$ of the capacitance $C_0$ and the charge
relaxation resistance $R_q$ based on the action we just derived.
We are interested in the response function $K (\omega)$, see Eq.~\eqref{linear} in the main text.
It is given in imaginary time by $K (\tau) = G(\tau)/\pi^2$ where we define the Green's function
\begin{equation}
G (\tau) =   \langle T_\tau \phi_0 (\tau) \phi_0 (0) \rangle,
\end{equation}
and the mean value $\langle \ldots \rangle$ is taken with respect to the full action $S_0 + S_{BS}$.
The perturbation calculation goes as follows: we expand the Green's function $G(\tau)$ 
in $S_{BS}$ and average with respect to the quadratic action $S_0$.
The result is finally analytically continued to real frequencies, $i \omega_n \to \omega+i 0^+$
for the retarded response function.
The lowest order is $G(\tau) = G_0 (\tau)$ with, for real frequencies,
\begin{equation}\label{expg0}
G_0 (\omega) = \frac{\pi^2}{2 E_c} \frac{1}{ 1  - \frac{i \omega \pi /E_c}{1-e^{2 i \omega L/v_F}}}.
\end{equation}
The same result is obtained for a large cavity where the exponential term is simply removed.
Eq.~\eqref{expg0} can be formally developed to first order in $\omega$,
\begin{equation}\label{exnoninter}
G_0 (\omega) = \frac{\pi^2}{2 E_c} \left( A + i \frac{\omega \pi}{E_c} B \right),
\end{equation}
with (i) $A^{-1} =1+\Delta/2 E_c$,
$B^{-1} = 2 (1+\Delta/2 E_c)^2$ for a small cavity and, (ii) $A= 1$, $B=1$ for a large cavity.
We now proceed with the expansion in $S_{BS}$ or $r$, the dimensionless strength of interface backscattering. 
For comparison, the transparency is $D = 1-r^2$ for $r \ll 1$.
The mean values $\langle \ldots \rangle_0$ 
with  the quadratic action $S_0$ are obtained by using variants of the general formula
\begin{equation}
\langle \phi_0 (\tau_1) \phi_0 (\tau_2) e^{2 i \phi_0 (\tau_3)} \rangle_0 
= \left( 2 i \langle \phi_0 (\tau_1) \phi_0 (\tau_3) \rangle_0 \right)
\left( 2 i \langle \phi_0 (\tau_2) \phi_0 (\tau_3) \rangle_0 \right) 
e^{-2 \langle \phi_0 (\tau_3) \phi_0 (\tau_3) \rangle_0}.
\end{equation}

The expansion to second order in $r$ then takes the form
\begin{equation}\label{corr-greens}
 G(i \omega_n)   = G_0 (i \omega_n) + 2 \bar{r}_0 \frac{[ G_0 (i \omega_n) ]^2}{A}
\left( e^{2 i \pi \tilde{N}_0} + e^{-2 i \pi \tilde{N}_0} \right)
-  \bar{r}_0^2 \frac{[ G_0 (i \omega_n) ]^2}{A^2}  \left[ 
\left( e^{4 i \pi \tilde{N}_0} + e^{-4 i \pi \tilde{N}_0} \right) F_+ (i \omega_n)
- 2 F_- (i \omega_n) \right],
\end{equation}
where the following couple of integrals has been defined 
\begin{equation}\label{setint}
F_{\pm} (i \omega_n)  =  \int_{-\beta/2}^{\beta/2} d \tau 
\left(e^{\mp 4 G_0 (\tau)}-1 \right)
\left(e^{-i \omega_n \tau} \pm 1 \right),
\end{equation}
and the renormalized expansion parameter reads
\begin{equation}\label{r0}
\tilde{r}_0  = \frac{\pi^2}{2 E_c} \, \frac{v_F r}{\pi a} \, A \, e^{-2 G_0 (\tau=0)}.
\end{equation}

A combination of complex plane contour deformations, essentially Wick's rotations,
allows to rewrite these expressions in a form that will be more convenient for
analytical continuation and low frequency expansion. It is worth discussing these
contour transformations in some details.
We start with the non-interacting Green's function $G_0 (\tau)$ taken in the 
zero temperature limit,
\begin{equation}
G_0 (\tau) = 2 T \sum_{n>0} G_0(i \omega_n) \cos ( \omega_n \tau)
= {\rm Re} \left( \int_0^{\omega_D} \frac{d \omega}{\pi} \, 
 G_0(i \omega) e^{-i \omega \tau} \right),
\end{equation}
where $ G_0(i \omega_n)$ is Eq.~\eqref{expg0} with $\omega = i \omega_n$.
A contour rotation $\omega = - i \omega'$ (later renamed $\omega$)
is performed leading to
\begin{equation}\label{noninter}
G_0 (\tau) = \int_0^{\omega_D} d \omega \, {\rm Im} 
\left(\frac{G_0 (\omega)}{\pi} \right) e^{-\omega \tau}.
\end{equation}
Note that a small contribution coming from frequencies 
larger than the high energy cutoff 
$\omega_D = v_F e^{-C}/a$ has been discarded in deriving Eq.~\eqref{noninter}.
The expression~\eqref{noninter} obtained for $G_0 (\tau)$ allows a trivial analytical
continuation to the complex plane for ${\rm Re} \, \tau \ge 0$.
For example, one has
\begin{equation}\label{noninter2}
G_0 (-i t) = \int_0^{\omega_D} d \omega \, {\rm Im} 
\left(\frac{G_0 (\omega)}{\pi} \right) e^{i \omega t}.
\end{equation}

Next we consider the set of integrals Eq.~\eqref{setint}, taken  at zero temperature,
namely
\begin{equation}
F_{\pm} (i \omega_n)  =   2 {\rm Re} \left( \int_0^{+\infty}   d \tau 
\left(e^{\mp 4 G_0 (\tau)}-1 \right)
\left(e^{-i \omega_n \tau} \pm 1 \right) \right),
\end{equation}
appearing inside the brackets of  Eq.~\eqref{corr-greens}.
Again a Wick's rotation $\tau = - i t$ is applied to these integrals
with the result
\begin{equation}\label{Fpm}
F_{\pm} (i \omega_n)  =  2  \int_0^{+\infty}   d t
\, \, {\rm Im} \left(e^{\mp 4 G_0 (-i t)} \right)
\left(e^{- \omega_n t} \pm 1 \right).
\end{equation}
The standard analytical continuation $i \omega_n \to \omega + i 0^+$ is
then straightforward to implement and Eq.~\eqref{corr-greens} transforms
into
\begin{equation}\label{corr-greens2}
 G(\omega)   = G_0 (\omega) + 2 \bar{r}_0 \frac{[ G_0 (\omega) ]^2}{A}
\left( e^{2 i \pi \tilde{N}_0} + e^{-2 i \pi \tilde{N}_0} \right)
-  \bar{r}_0^2 \frac{[ G_0 (\omega) ]^2}{A^2}  \left[ 
\left( e^{4 i \pi \tilde{N}_0} + e^{-4 i \pi \tilde{N}_0} \right) F_+ (\omega)
- 2 F_- (\omega) \right].
\end{equation}

The last step of this calculation is to expand the Green's function
$G(\omega)$ to first order in the frequency $\omega$.
The expansion of the non-interacting Green's function  $G_0 (\omega)$
has already been given in Eq.~\eqref{exnoninter}.
It is a fascinating fact  that the first order term in the expansion of $F_{\pm} (\omega)$
can be calculated exactly.
Let us note $F_{\pm} (\omega) = F_{\pm} (0) \mp i \omega F_1$ with 
\begin{equation}
F_1   =  \mp 2  \int_0^{+\infty}   d t \, t \,
{\rm Im} \left(e^{\mp 4 G_0 (-i t)} \right).
\end{equation}
The imaginary part inside the integral allows to add arbitrary 
real terms inside the parantheses. 
Therefore, using the expression~\eqref{noninter2} for the non-interacting
Green's function, we write
\begin{equation}\label{ima}
F_1   =  2 {\rm Im} \left[   \int_0^{+\infty}   d t \, t \,
 \left( e^{\mp 4 \int_0^{\omega_D} d \omega \, {\rm Im} 
\left(\frac{G_0 (\omega)}{\pi} \right) e^{\mp i \omega t} } - 1 - \frac{\pi^3}{2 E_c^2} 
\frac{ 4 B / \pi}{1+t^2}  \right) \right].
\end{equation}
The last two terms have been chosen to cancel the leading asymptotical
behaviour of 
\begin{equation}
{\rm Exp} \left[ \mp 4 \int_0^{\omega_D} d \omega \, {\rm Im} 
\left(\frac{G_0 (\omega)}{\pi} \right) e^{\mp i \omega t} \right],
\end{equation}
at large $t$. It involves small $\omega$ and therefore 
the coefficient $B$, see Eq.~\eqref{exnoninter}, when the imaginary part
is taken.
The large $t$ asymptotical behaviour of the integrand in Eq.~\eqref{ima}
is faster than $1/t$ so that the integral over $t$ can be rotated by $\mp \pi/2$
with a pole at $t = \mp i$.
The imaginary part in Eq.~\eqref{ima} cancels all terms but the pole's contribution.
The final result is simply
\begin{equation}
F_1 = \frac{2 \pi^3}{E_c^2} B,
\end{equation}
and it does not depend on the sign $\pm$ as anticipated in our notation.
Inserting this result into Eq.~\eqref{corr-greens2}, we are able expand to first order in $\omega$.
The low frequency expansion for $K (\omega) = G (\omega)/\pi^2$ then reads
\begin{equation}\label{lowfreq}
\begin{split}
& K (\omega) = \frac{ A}{2 E_c} \left[ 1 + 2 \bar{r}_0 
\left( e^{2 i \pi \tilde{N}_0} + e^{-2 i \pi \tilde{N}_0} \right)
+ \frac{\bar{r}_0^2}{A} \, F_0 \left( e^{4 i \pi \tilde{N}_0} + e^{-4 i \pi \tilde{N}_0} \right) \right] \\[2mm]
&+ \frac{i \omega \pi}{E_c}  \frac{B}{2 E_c} \Bigg[
1 + 4 \bar{r}_0 
\left( e^{2 i \pi \tilde{N}_0} + e^{-2 i \pi \tilde{N}_0} \right) 
+ \frac{2 \bar{r}_0^2}{A} \, F_0 \left( e^{4 i \pi \tilde{N}_0} + e^{-4 i \pi \tilde{N}_0} \right) 
  + 4 \bar{r}_0^2 \left( e^{4 i \pi \tilde{N}_0} + e^{-4 i \pi \tilde{N}_0} +2 \right) \Bigg],
\end{split}
\end{equation}
where $F_0 \equiv - F_+ (0)$. Using this result in  Eq.~\eqref{linear}
of the main text, we readily demonstrate from a comparison with the classical RC circuit, 
Eq.~\eqref{charge} in the main text, that the charge relaxation resistance is
\begin{equation}
R_q = \frac{h}{e^2} \, \frac{B}{A^2},
\end{equation}
regardless of the backscattering $r$. It is worth emphasizing 
that this result was proved 
for general values of the coefficients $A$ and $B$ in Eq.~\eqref{exnoninter}.

The capacitance is obtained from the zero frequency result, $C_0 = e^2 K(0)$.
We first check that $e^2 A/(2 E_c) = C_\mu$. Next, we use the explicit expression for
the non-interacting Green's function~\eqref{expg0} in Eq.~\eqref{noninter}
at $\tau = 0$, and insert the result into Eq.~\eqref{r0}
to compute the parameter $\tilde{r}_0$. We find
\begin{equation}
\tilde{r}_0 = \frac{r e^C}{\pi} \, \frac{\pi}{2} \, g_1 \left( \frac{\Delta}{E_c} \right),
\end{equation}
 where $C = 0.577\ldots$ is the Euler constant and the function
\begin{equation}
g_1 (\alpha) = \frac{2}{1+\alpha/2} \, {\rm Exp} \left[
\int_0^{+\infty} d u \left( \frac{1}{1+u} - \frac{2 u}{u^2 + [1+ u \cot (2 u /\alpha)]^2}
\right) \right],
\end{equation}
has been defined. The limiting values are $g_1(0) = 1$ and $g_1 (\infty)= e^{-C}$.
A second function is defined $g_2 (\Delta/E_c) = (\pi/8) F_0/A$ ($F_0 = - F_+(0)$).
It can be computed from Eq.~\eqref{expg0},   Eq.~\eqref{noninter2} and Eq.~\eqref{Fpm} 
for $i \omega_n=0$ with the outcome
\begin{equation}
g_2 (\alpha) = - (1+\alpha/2) \, \int_0^{+\infty} d x \,
{\rm Im} \left[ {\rm Exp} \left( -4 \int_0^{+\infty} d y 
\frac{y e^{i x y}}{y^2+[2+y \cot(y/\alpha)]^2} \right) \right].
\end{equation}
The limiting values are $g_2(0) \simeq 1.86$ and  $g_2(\infty) = \pi/2$.
The final result is given Eq.~\eqref{resultcapa} in the main text.




\end{document}